\documentclass[manuscript]{acmart}

\AtBeginDocument{%
  \providecommand\BibTeX{{%
    \normalfont B\kern-0.5em{\scshape i\kern-0.25em b}\kern-0.8em\TeX}}}

\setcopyright{acmcopyright}
\copyrightyear{2022}
\acmYear{2022}
\acmDOI{XXXXXXX.XXXXXXX}

\acmJournal{JACM}
\acmVolume{37}
\acmNumber{4}
\acmArticle{111}
\acmMonth{8}

\acmPrice{15.00}
\acmISBN{978-1-4503-XXXX-X/18/06}

\begin{document}

\title{Bayesian Prior Learning via Neural Networks for Next-item Recommendation}


\author{Manoj Reddy Dareddy}
\email{mdareddy@cs.ucla.edu}
\affiliation{
  \institution{University of California Los Angeles}
  \city{Los Angeles}
  \state{California}
  \country{USA}
}

\author{Zijun Xue}
\email{xuezijun@cs.ucla.edu}
\affiliation{
  \institution{University of California Los Angeles}
  \city{Los Angeles}
  \state{California}
  \country{USA}
}

\author{Nicholas Lin}
\email{nlin37@ucla.edu}
\affiliation{
  \institution{University of California Los Angeles}
  \city{Los Angeles}
  \state{California}
  \country{USA}
}

\author{Junghoo Cho}
\email{cho@cs.ucla.edu}
\affiliation{
  \institution{University of California Los Angeles}
  \city{Los Angeles}
  \state{California}
  \country{USA}
}

\renewcommand{\shortauthors}{Dareddy et al.}

\begin{abstract}
  Next-item prediction is a a popular problem in the recommender systems domain. As the name suggests, the task is to recommend subsequent items that an user would be interested in given contextual information and historical interaction data. In our paper, we model a general notion of context via a sequence of item interactions. We model the next item prediction problem using the Bayesian framework and capture the probability of appearance of a sequence through the posterior mean of the Beta distribution. We train two neural networks to accurately predict the alpha \& beta parameter values of the Beta distribution. Our novel approach of combining black-box style neural networks, known to be suitable for function approximation with Bayesian estimation methods have resulted in an innovative method that outperforms various state-of-the-art baselines. We demonstrate the effectiveness of our method in the song recommendation domain using the Spotify playlist continuation dataset.
\end{abstract}



\maketitle

\section{Introduction}

Next-item prediction is a classic problem in the domain of Natural Language Processing (NLP) and Recommender Systems. In the NLP domain, it is used for the purpose of language modeling \cite{jozefowicz2016exploring} whereby the task is given a set of tokens, predict the most likely next token. In natural language, words can be represented by tokens and accurately predicting the next token helps us understand the intricacies of a particular language.
In Recommender Systems, we maybe given access to historical transactions by a set of users such as products purchased, movies watched etc. The next item prediction problem in this domain \cite{adomavicius2005toward} boils down to the task of predicting the next item/movie a user is most likely to purchase/watch.

\begin{figure*}
\centering
\includegraphics[scale=0.3]{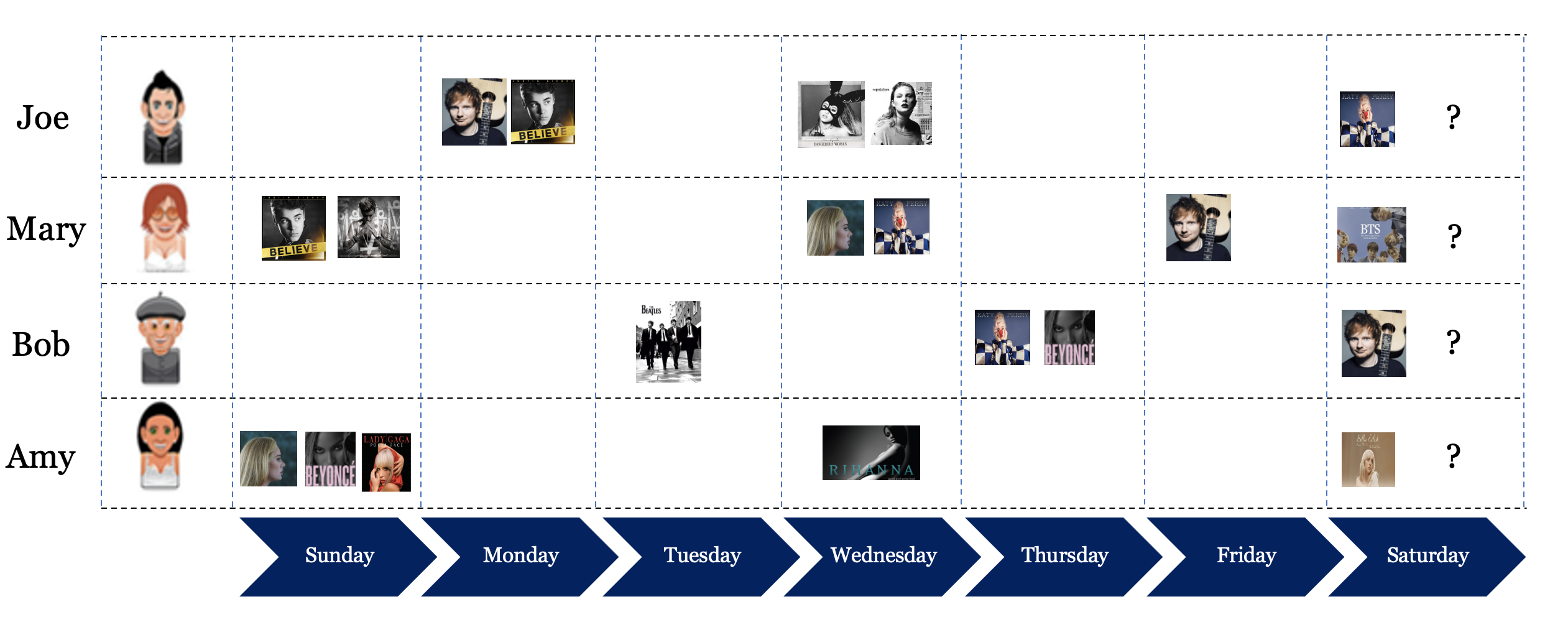}
\caption{Status Quo of Next Item Recommendation}
\label{bayesian_statusquo}
\end{figure*}

As illustrated in Figure \ref{bayesian_statusquo}, existing next-item recommender systems track user behavior over time, and utilize this information to provide recommendations of items they are most likely to consume in a session. Recommender Systems that track the entire user history in general are able to provide high-quality recommendations since they do so at scale across millions of users. The drawback of such systems, is that the accuracy comes at the expense of user privacy, since each individual user's activity is being logged by the system. Moreover, users do not have control over how their data is being utilized on the server which results in the loss of their privacy.

Various approaches have been proposed in the literature for the next-item prediction problem. Majority of these techniques rely on an unsupervised training dataset which captures previous interactions. For instance, in the NLP domain it can refer to a large corpus of unstructured documents, whereby sentences are represented as a sequence of tokens/words. In the recommender systems domain it can refer to sequences of historical purchases of items that are organized in sessions. 


Majority of the existing approaches tackle the next item problem using statistical techniques by analyzing for patterns across the provided training dataset. By sifting through large amounts of training data, these models are able to fairly accurately predict the next item at test time given a seed context.

The proposed approaches fall under the following two categories: black-box neural network and hand-crafted approaches. The former set of approaches employ the use of neural networks which are known to have high representational power and perform implicit feature engineering. Among the latter miscellaneous hand-crafted approaches, Bayesian methods are known to be intuitive and easy to understand/debug. Each approach has their own set of pros and cons. Neural network approaches perform very well in practice but have the drawback of being considered as a black-box, i.e. provides limited understanding of the underlying prediction mechanism. Bayesian approaches on the other hand are based on the Bayes rule \cite{stone2013bayes}, a fundamental concept in probability making it intuitive and easy to understand. The drawback of a pure Bayesian approach is that it does not perform as well as the neural network approaches in the task of next-item prediction.


In this work, our main contribution is as follows:
\begin{itemize}
    \item We propose a Bayesian framework with prior information learnt using neural networks for the next-item prediction.
    
    \item We introduce the confidence of observed data using the Beta distribution and utilize a Siamese network to estimate the Beta model parameters.
    
    \item We demonstrated the performance improvements our method against existing state-of-the-art approaches.
    
\end{itemize}

The rest of the paper is organized as follows. In section 2, we describe the related work existing in the literature. In section 3, we describe our approach in detail along with our motivation and intuitions. In section 4, we describe the experimental setup and detailed training process. In section 5, we present our results in comparison with various state-of-the-art baselines. In section 6, we analyze our results and provide some insights into our understanding of the problem. In section 7, we propose further directions that can be pursued to expand this research area. In section 8, we conclude by highlighting our novel contribution to the next-item prediction problem.


\section{Related Work}

The related work in this area consists of two main categories, namely: neighborhood based and neural network based approaches. There exist other types of approaches which we describe later in this section.

Neural network based approaches mostly reply on the sequence based models using the Recurrent Neural Network architecture. Li et al. \cite{li2018learning} propose an item embedding method based on an aggregation of the user interactions. They then use a contextual LSTM neural network architecture to train on two real-world datasets in order to predict the next-item. Wu et al. \cite{wu2019atm} expand on attention mechanism idea proposed by Vaswani et al. \cite{vaswani2017attention} which is known to be efficient and also work well in the natural language processing domain. They were able to demonstrate the superiority of the attention mechanism over the LSTM approach.

In addition to above mentioned approaches, there exist unconventional yet effective methods to recommendation under different settings. For example, Dareddy et al. \cite{dareddy2019motif2vec} propose the use of motifs in heterogeneous networks for the task of link recommendation.
Similarly, Mishra et al. \cite{mishra2016bottom} proposed a novel next-item prediction solution tailored towards job recommendation whereby users were matched with jobs based on a variety of context features.


\section{Our Approach}

\subsection{Intuition}

Our problem can be formulated as follows:  $\mathcal{I} =\{a, b, c ...\}$ represents a set of items, $\mathcal{S} = (S_1, S_2, ... S_n)$ represents a user interaction sequence with $S_i \in \mathcal{I}$. We observe many user sequences and the goal is given a length k subsequence C of S, $C=(S_i, S_{i+1}, ... S_{i+k-1}) \subset S$, predict the next item in $C$ i.e. to accurately estimate the following probability for a specific item $a$:
\begin{equation}
P(S_{i+k}=a | (S_i, S_{i+1}, ... S_{i+k-1}))
\end{equation}
An accurate estimation of the above probability requires analyzing our observed training data.
The intuition is that given a longer context information, perhaps we can more accurately compute the above posterior probability value.

Assume we are given a context of 3 items, namely $(a, b, c)$ and we are interested in calculating the next-item probability values for 2 items namely: $x$ and $y$. Hence we are interested in computing the following values: 
\begin{center}
$P(x|(a,b,c))$ \& $P(y|(a,b,c))$
\end{center}
Consider two scenarios for the frequency of n-gram values in our training dataset. Scenario I is described as below:

\begin{table}[H]
\centering
\begin{tabular}{|c|c|c|c|c|c|c|c|c|}
\hline
Prefix    & \multicolumn{2}{c|}{abc} & \multicolumn{2}{c|}{bc} & \multicolumn{2}{c|}{c} & \multicolumn{2}{c|}{-} \\ \hline
Sequence  & abcx        & abcy       & bcx        & bcy        & cx         & cy        & x         & y          \\ \hline
Frequency & 100         & 50         & 200        & 100        & 300        & 150       & 500       & 1000       \\ \hline
\end{tabular}
\caption{\label{tab:scenario-1}Scenario I Frequency Values}
\end{table}

Based on the frequency values in Table \ref{tab:scenario-1}, we observe that in general item y is more popular than item x since their frequencies in the train dataset are 1000 and 500 respectively. But under various contexts \{(c), (b, c) \& (a, b, c)\}, we see that item x appears more than item y. This indicates that item x would be a better choice than item y for next item prediction given context (a, b, c).


Consider another Scenario II whereby the frequency information of sequences is as follows:

\begin{table}[H]
\centering
\begin{tabular}{|c|c|c|c|c|c|c|c|c|}
\hline
Prefix    & \multicolumn{2}{c|}{abc} & \multicolumn{2}{c|}{bc} & \multicolumn{2}{c|}{c} & \multicolumn{2}{c|}{-} \\ \hline
Sequence  & abcx        & abcy       & bcx        & bcy        & cx        & cy         & x         & y          \\ \hline
Frequency & 2           & 1          & 5          & 500        & 50        & 750        & 500       & 1000       \\ \hline
\end{tabular}
\caption{\label{tab:scenario-2}Scenario II Frequency Values}
\end{table}

Similar to Scenario I, in general item y is popular than item x (same frequencies as before). In contrast to the previous scenario, for the context (a, b, c), item x seems to be a better choice than item y, due to the frequencies being 2 and 1 respectively. Although item x appears twice as many times as item y under the context (a, b, c) it is not a reliable signal since the raw frequency values are very low. This results in a low confidence of accuracy for likelihood estimation under the context (a, b, c). Instead if we rely on a smaller context (b, c), we observe that item y is clearly a better choice than item x since it co-occurs 500 compared to 5 times respectively. We observe the same phenomenon under the context of single item (c). Overall, the observations under smaller contexts can be considered to be more confident in recommending item y as the better next item recommendation under this scenario.


\begin{figure}[!h]
\centering
\includegraphics[scale=1.0]{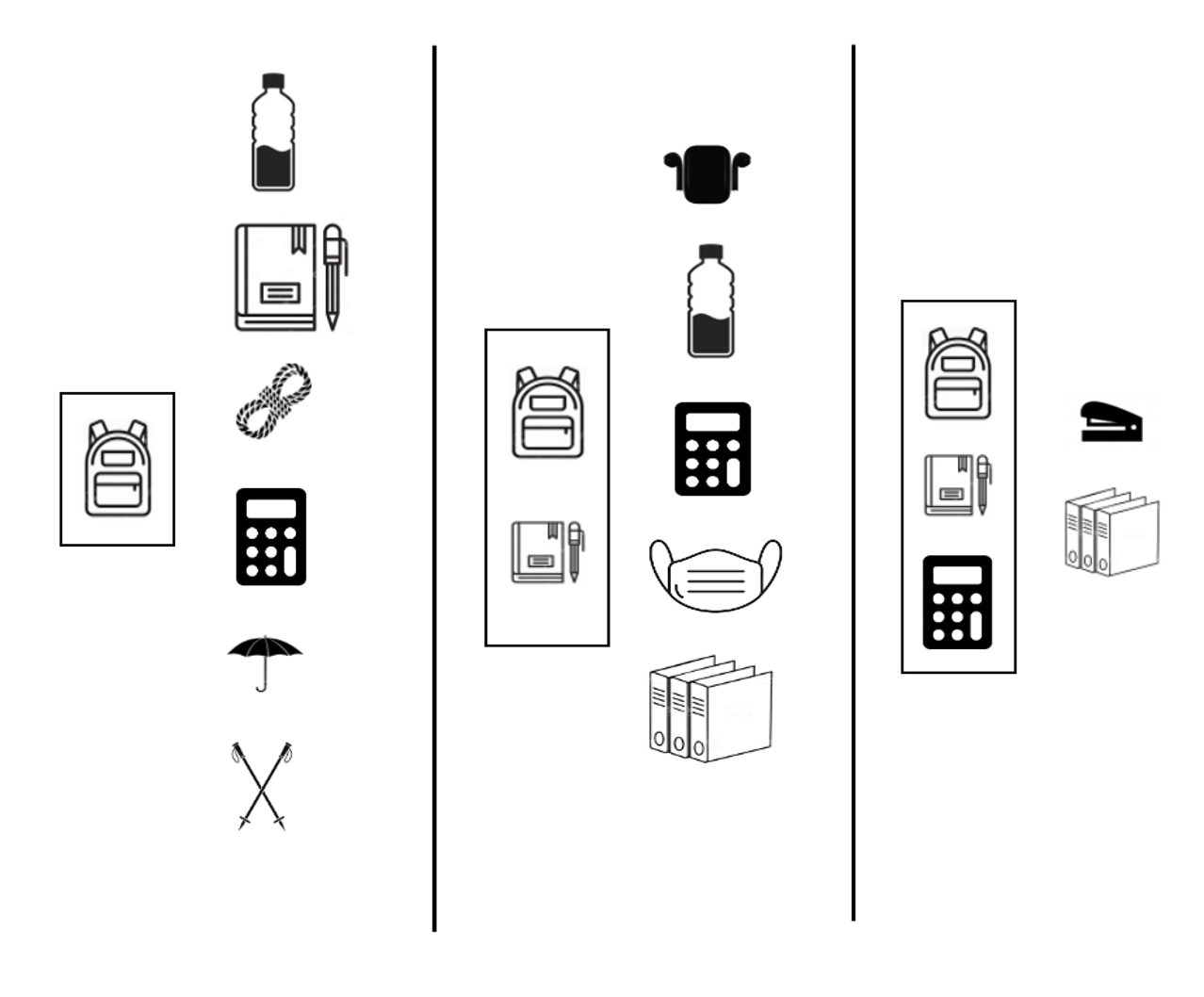}
\caption{Longer context leads to better prediction accuracy}
\label{bayesian_intuition1}
\end{figure}

Consider a scenario as shown in Figure \ref{bayesian_intuition1} whereby we are tracking the items a user has purchased in a given session. Given that the user has just purchased a bagpack, the systems recommends items that are related to the initial purchase, such as: waterbottle, diary, hiking ropes, calculator, umbrella and trekking poles. These items would be considered good recommendations since we are not sure whether the user is buying a bagpack for school, travel or hiking purposes. In the next column, the user purchases a diary after the bagpack purchase, indicating that the user is likely to be interested in purchasing items that are school-related. We can be more confident about the items to recommend once we know that the user has purchased a bagpack followed by diary and a calculator. In the case, we can be confident that the user is interested in purchasing stationary supplies, hence items such as stapler, folders etc. would be good recommendations. The main intuition here is that the more context information we have, the better we can predict the next item, simply because we have access to more information from the context.

\begin{figure*}
\centering
\includegraphics[scale=0.8]{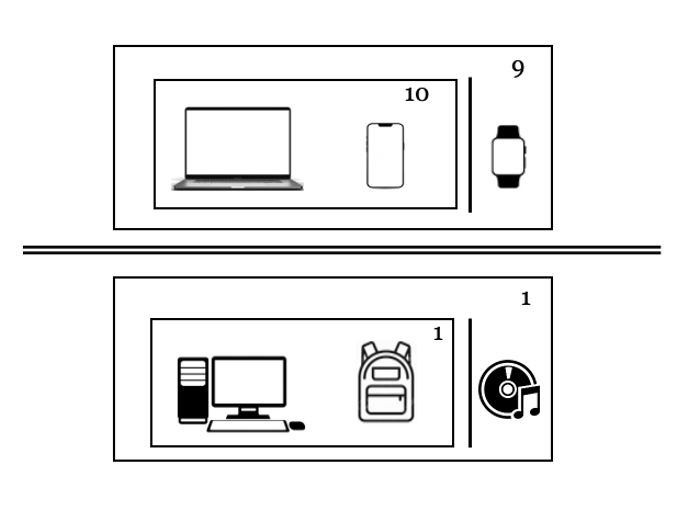}
\caption{Higher frequency indicates higher confidence}
\label{bayesian_intuition2}
\end{figure*}

Consider another scenario whereby we are analyzing the purchases made by users in a system. Let's say we observe that out of 10 purchase sequences of a laptop followed by a mobile, we observe that users have purchased a watch 9 times. This implies that it's likely that given a sequence of laptop and mobile purchase, the user is likely to buy a watch about 90\% of the transactions. In the second scenario, given that we observe only one transaction across the entire transactions log of the following of items: computer, bagpack and a music label. In the future, given that the user has purchased a PC followed by a bagpack, then the next item to recommend based on our history would be the music label with a probability of 100\%. In reality, the music label is not a good recommendation since we observe the entire transaction only once. The issue here is that of data sparsity. Given a limited set of observations, it is inherently difficult to estimate the likelihood probability. In order to mitigate the challenge of data sparsity, we capture our belief of the probabilities using the posterior distribution for every item. Intuitively, instead of modeling the point estimates based on ratios obtained from historical transactions, we ought to model the distribution of the point estimate based on the raw frequency counts of sequences observed. In particular we use the Beta distribution.

\begin{figure*}
\centering
\includegraphics[scale=1.0]{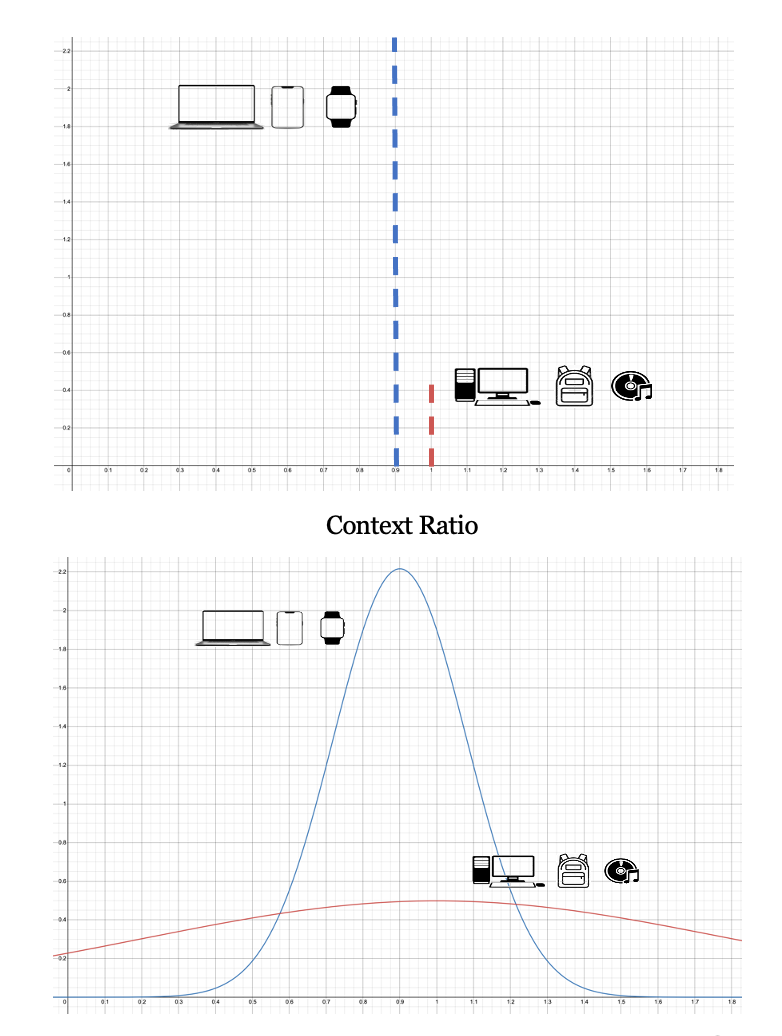}
\caption{Modeling the posterior distribution instead of raw point estimate}
\label{bayesian_intuition3}
\end{figure*}

Beta distribution is well known to be a good prior for various distributions such as Binomial, Normal and is widely used in practice as well \cite{armagan2011generalized}. We chose the Beta distribution since it is able to effectively model binary feedback. The Beta distribution consists of 2 parameters, namely: alpha ($\alpha$) \& beta ($\beta$). The parameter alpha captures the positive count whereas the beta distribution captures the corresponding negative count. The mean of a Beta distribution summarizes the overall binary feedback.

Given a set of $\alpha$ \& $\beta$ values, the mean of the Beta distribution is as follows:
\begin{equation}
    \mu(\alpha, \beta) = \frac{\alpha}{\alpha+\beta}
\end{equation}

The mean of the Beta distribution can provide insights into the overall likelihood of observing a given sequence.

Our goal is to incorporate the counts of various context frequencies so that the neural network can learn the most appropriate $\alpha$ \& $\beta$ values from our training data. The mean based on alpha \& beta values will estimate our likelihood for the probabilities of various context, item pairs.

Without loss of generality, we restrict ourselves to a context size of four items/songs. Given a sequence of 4 songs a, b, c, d, we would like to be able to accurately estimate the posterior probability of these 4 songs in order. By doing so, given a test seed playlist of 3 songs and a set of candidates, we can rank the candidates based on the posterior mean and recommend the one with the highest value.

We aim to learn 2 separate functions to accurately predict the alpha and beta values. The idea to ingest frequency counts from the training datasets and learn to accurately predict the alpha and beta values.

Given we observe a sequence of 4 songs ‘a, b, c, d’ in our dataset, we initially collect the following counts/frequency values:

\textbf{Positive Counts:}
\begin{itemize}
    \item \#(d)
    \item \#(cd)
    \item \#(bcd)
    \item \#(abcd) 
\end{itemize}

\textbf{Negative Counts:}
\begin{itemize}
    \item \#($\sim$d)
    \item \#(c$\sim$d)
    \item \#(bc$\sim$d)
    \item \#(abc$\sim$d)
\end{itemize}

The above values are used as input to the functions that predict the alpha \& beta parameter values. We aim to learn these functions in a \textit{contrastive} learning method. For an observed sequence of 4 items in the training data, which we denote as a positive example, we would like to use a negative example so that the functions can learn to maximize the posterior Beta distribution mean between the two.


Our methodology for obtaining a negative example given an observed positive sequence of ‘a, b, c, d’ is as follows:

\begin{itemize}
    \item If ‘d’ is the most popular song given the prefix ‘a, b, c’ then we select ‘a, b, c, e’ to be the negative example whereby ‘e’ is the next most popular song after ‘d’.
    
    \item If ‘d’ is not the most popular song given the prefix ‘a, b, c’ then we select ‘a, b, c, e’ to be the negative sequence whereby ‘e’ is the most popular song given the prefix ‘a, b, c’.
\end{itemize}

The intuition is that we would like to maximize the difference between the posterior means of the positive and negative samples which are computed using the alpha and beta functions. The alpha and beta functions are learning to predict the accurate alpha and beta parameter values for the positive and negative examples based on their frequency counts as mentioned in the above table.

Without loss of generality, given positive example ‘a, b, c, d’ and the corresponding negative example ‘a, b, c, e’ which is a single training instance to the overall neural network, we initially compute the following four values:


$    \alpha_+ = f_\alpha(positive\_example\_counts)$
    
$    \beta_+ = f_\beta(positive\_example\_counts) $
    
$    \alpha_- = f_\alpha(negative\_example\_counts) $
    
$     \beta_- = f_\beta(negative\_example\_counts)$

The mean of the positive example becomes: 
\begin{equation}
Positive\_mean =  \frac{\alpha_+}{\alpha_+ + \beta_+}
\end{equation}

Correspondingly, the mean of the negative example becomes:

\begin{equation}
Negative\_mean = \frac{\alpha_-}{\alpha_- + \beta_-}
\end{equation}

The goal is to maximize the following value: 
\begin{equation}
    ( Positive\_mean - Negative\_mean )
\end{equation}

\begin{figure*}
\centering
\includegraphics[scale=1.0]{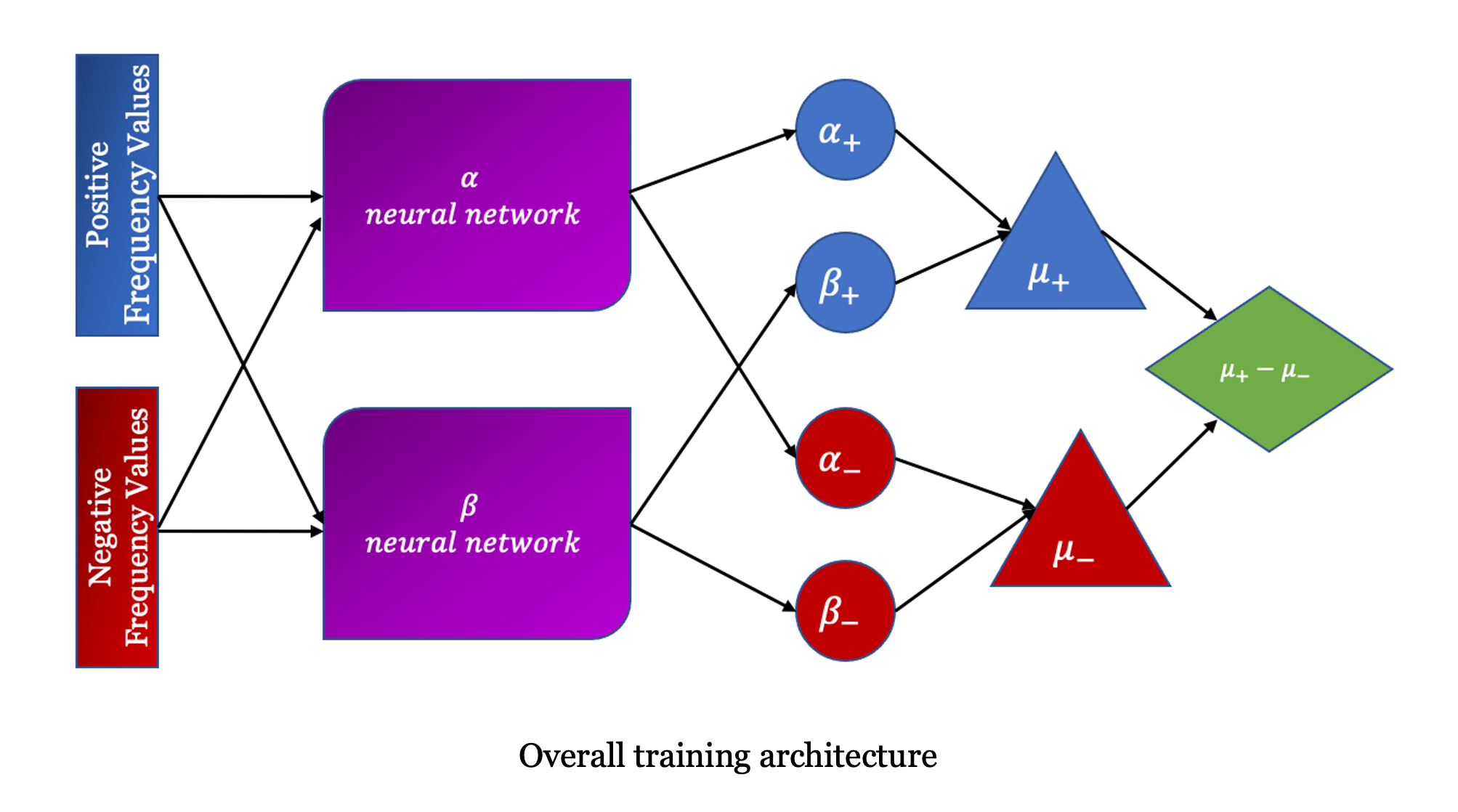}
\caption{Proposed Neural Network Architecture}
\end{figure*}

\begin{figure*}
\centering
\includegraphics[scale=1.2]{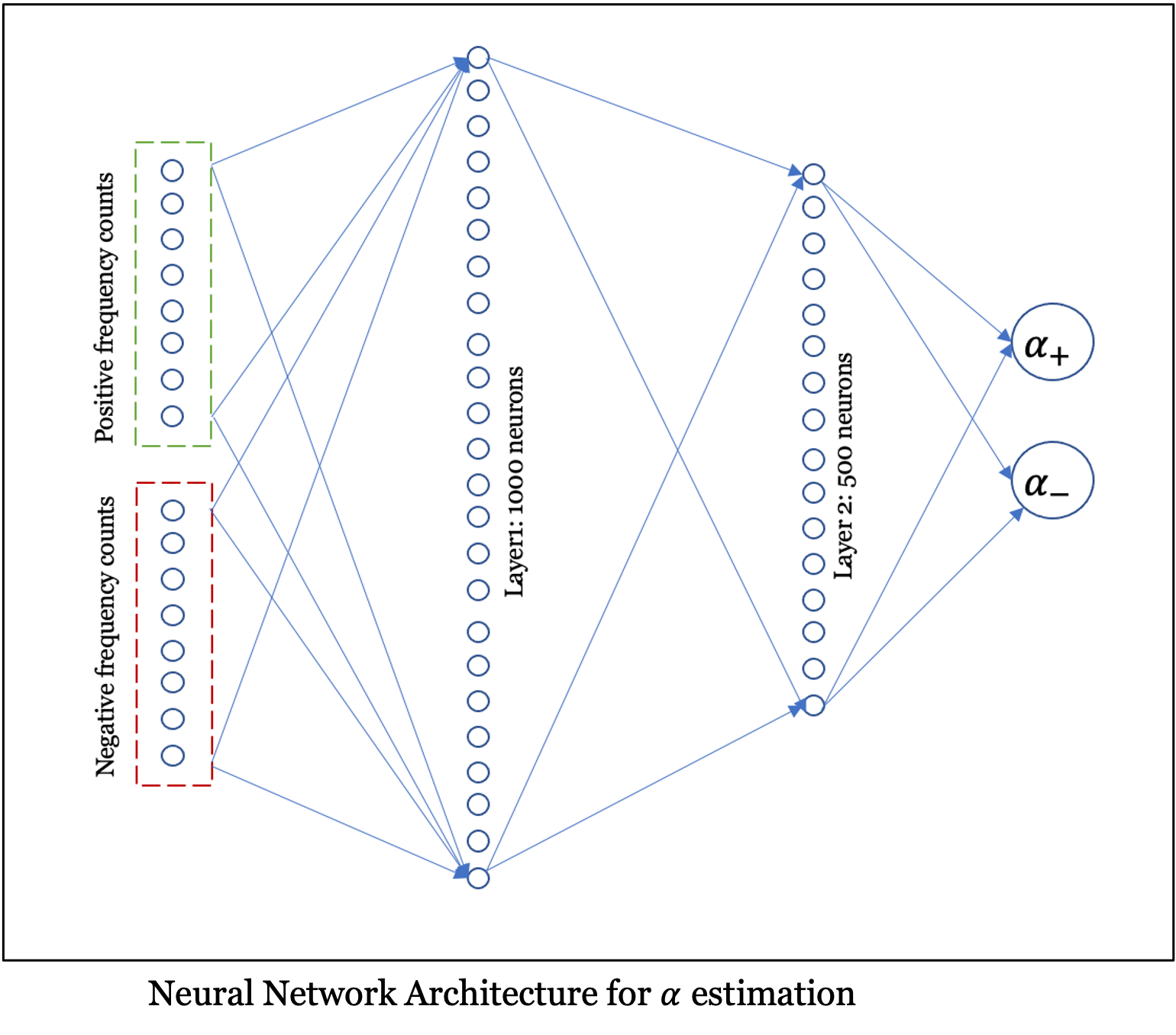}
\caption{Detailed Alpha/Beta Function}
\end{figure*}


We employ neural networks as the functions for the estimation of alpha and beta values as shown in the above figure.
Neural networks are known to be very effective at the task of function approximation. We use the feed-forward fully connected neural network both our alpha and beta functions. Each of them consist of an input layer with 8 neurons corresponding to the frequency counts, and two intermediate layers with 1000 and 500 neurons respectively.


\section{Experimental Setup}

\subsection{Dataset}

Spotify, a popular music streaming service, released a dataset for the RecSys 2018 Challenge \cite{chen2018recsys}. The dataset consists of a collection of playlists whereby, a playlist is an ordered sequence of songs. Playlists can be of varying lengths and a song can occur multiple times within a single playlist. The dataset is composed of 100,000 playlists. The distribution of playlist length demonstrates playlists tend to have less than 50 songs total. The most common playlist length is 20 songs, with its frequency being approximately 1.5\% of the training dataset. There are 686,685 unique songs among these playlists. Although songs can occur multiple times, 676,244 of the unique 686,685, a vast majority, occurred less than 100 times across the entire dataset. A more detailed overview of song distribution can be found in Table 1. 

\begin{figure}
\centering
\includegraphics[width=0.8\textwidth]{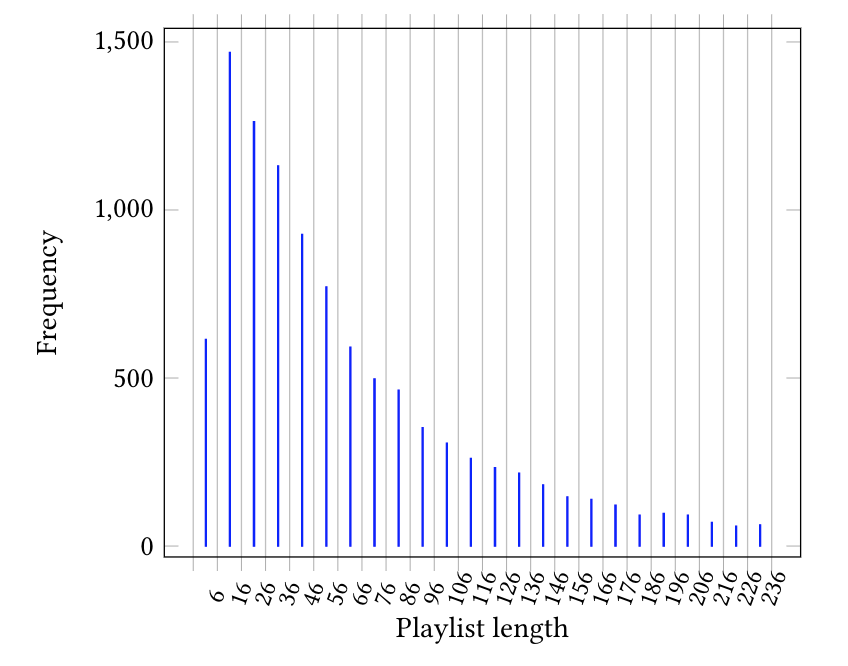}
\caption{Playlist Length Frequency Distribution}
\end{figure}

\begin{figure}
\centering
\includegraphics[width=0.8\textwidth]{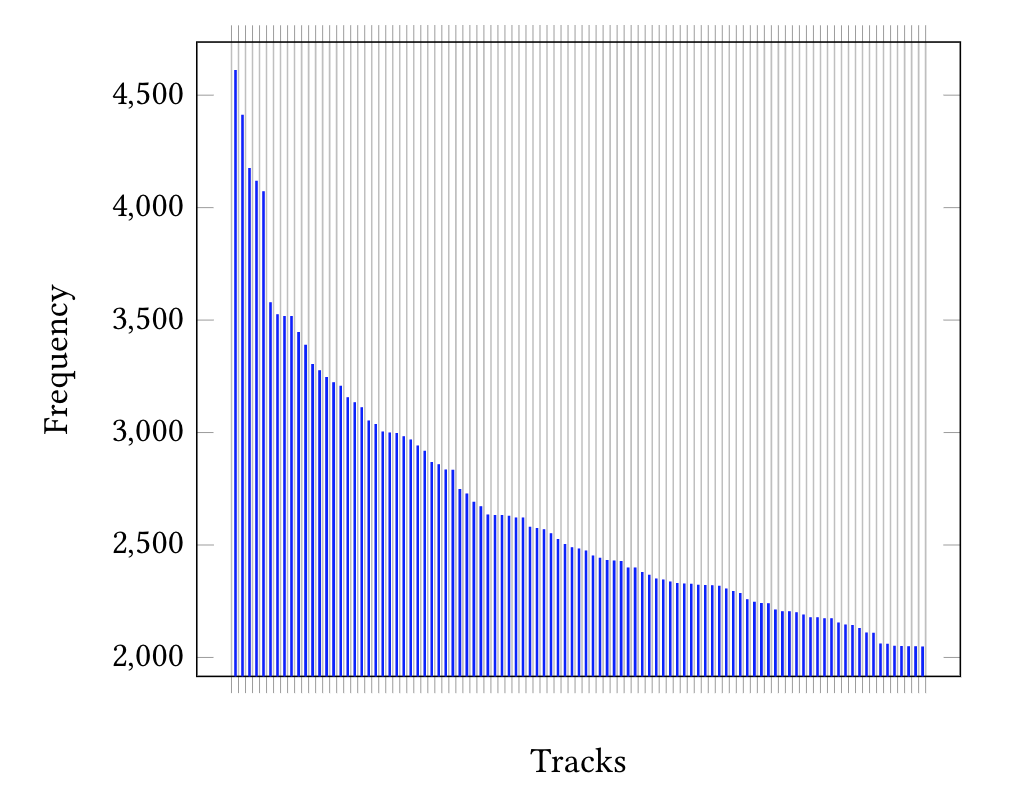}
\caption{Track frequency distribution across playlists}
\end{figure}

The alpha and beta functions are neural networks with the following layout:

8 -> 1000 -> 500 -> 1

The input layer is a vector of 8 scalar values which represent the various frequency counts described earlier for a given sequence ‘a, b, c, d’.
The 1000-neuron layer after the input layer has an exponential activation function.

The final output of the network is a scalar with an exponential activation function.

\subsection{Training Process}

Given a train file with a list of playlists, the first step involves collecting statistics of various gram data. We utilize a Trie datastructure to store the frequencies of 1,2,3 \& 4 gram song sequences across the playlists. The Trie \cite{bodon2003trie} datastructure enables efficient access of the counts via the prefix notation. Trie is a tree-based datastructure whereby a node can have multiple children and the edges represent an item. The node in a Trie captures a sequence of items and can be used to store relevant information, in our case the frequency of a particular sequence so far. The only bottleneck of Trie data structure is insertion time complexity. In our setting, we perform insertion only once and the Trie offers quick retrieval of frequency count information for various sequences of items. The benefit of using a Trie over a HashTable is that Trie takes advantage of the prefix structure of sequences and hence requires significantly less memory compared to a HashTable based implementation.

Once the Trie datastructure has been populated, we need to generate the training file which is the input to our neural network. For each playlist, we go over a sequence of songs with a sliding window of size 4. This window of 4 songs would be considered as a positive instance since we have observed it in our training dataset. Given our positive instance ['a', 'b', 'c', 'd'], we generate a corresponding negative instance with the same prefix, ['a', 'b', 'c', 'e']. The negative instance is chosen based on the prefix ['a', 'b', 'c']. Item 'e' is chosen to be the most popular item given the prefix ['a', 'b', 'c']. If 'd' itself is the most popular item then 'e' is chosen to be second-most popular item given the prefix after 'd'.

This positive and negative instance pair become one training example for the neural network. Each instance is represented using the 8 count statistics described earlier. We place the counts of the instance pair adjacent to each other these set of 16 values becomes 1 training instance.

By using a sliding window approach we are able to generate multiple training instances per playlist and store all of them in the training file.

The hinge loss is used since our goal is simply to maximize the difference between the posterior mean of the positive and its corresponding negative example. Hinge loss \cite{gentile1999linear} is defined as follows: 
\begin{equation}
    max(0, 1-y\_true*y\_pred)
\end{equation}

In our setting, the y\_true value is not irrelevant and is set to a fixed scalar value for all training examples.
The benefit of using hinge loss is that it only penalizes when the predicted value is opposite of the ground truth and loss remains 0 otherwise.
In our case, when the y\_pred value which is the difference between positive and negative instance is above 0, then it doesn't contribute to the loss.
Meanwhile, when our network predicts a negative value, the loss becomes (1-y\_true*y\_pred) which becomes a positive value. The exact loss value depends on the magnitude of the difference in the mean did the network predict.
Backpropagation \cite{chauvin1995backpropagation} ensures that weights across the network are modified in order to reduce the loss value inturn, maximizing the difference between posterior means of positive and negative examples.


\subsection{Candidate Generation}
To improve efficiency, the underlying frequency count information is stored in a Trie data structure. The frequency counts are dependent on the training dataset and once computed does not change. Hence, the Trie data structure allows for efficient retrieval count given a prefix which is essential operation for most of our computation.

During testing, given a set of candidates ‘A, B, C’ we generate a set of candidates as follows: 
Generate the set of items that appear followed by ‘A, B, C’, calling it S. We then sort the items in set S in decreasing order of frequency and then select a predefined top-k items.
We then rank the items in the candidate set based on our estimated posterior mean and select the item with the highest mean.

\subsection{Evaluation Metrics}

We evaluate our approach using the standard metrics used in next item prediction, namely:
\begin{itemize}
    \item Overall Accuracy

This metric computes the percentage of next item predictions by the model that have been accurate in the test set.

\item Recall@20

Recall is defined as the fraction of relevant items that have been retrieved. 

\begin{equation}
Recall@20 = \frac{|retrieved@20 \cap relevant@20|}{|relevant@20|}
\end{equation}

Recall@20 restricts the analysis to the first 20 relevant and retrieved items, whereby the order information is not taken into account. A higher Recall@20 value indicates greater overlap between the ground truth and next items predicted by the model.

\item Mean Reciprocal Rank

Mean Reciprocal Rank (MRR) measures the average reciprocal rank of the ground truth next item across users in the test set.

Intuitively, lower the rank value of the ground truth next item, higher the MRR value indicating better performance.

\end{itemize}

\subsection{Baselines}
To better understand the results of our approach in the domain of next-item prediction, we compared the results of several other popular approaches to our approach on our Spotify dataset. Not only do these baselines include popular approaches towards next-item prediction using modern machine learning models, but our baselines also include simplistic approaches such as recommending the most popular song. The details of each baseline used for comparison can be found below.

Our baseline is to use the most popular item observed from the training data given a prefix of 3 songs. This most popular baseline is easy to implement and surprisingly performs very well on the test data.

\subsubsection{Nearest Neighborhood}
As explained by Kelen et al., the nearest neighborhood approach performs quite well for playlist continuation\cite{kelen2018efficient} and was previously utilized on the Spotify dataset. In this approach, a playlist-track matrix is created and then utilized to create a playlist-based neighborhood model that is then used for playlist continuation. In the original implementation of the paper, the neighborhood model extends the playlist with 500 additional songs, but as a baseline we utilize the first song recommended as the sixth song given the first five seeds.

\subsubsection{Nearest Embedding}
In this approach, an embedding for every song in the dataset is computed. We employ techniques such as Word2Vec \cite{mikolov2013distributed}, which look at the context information across a sliding window in playlists to place songs in an multi-dimensional embedding space. 

In our baseline, we utilize a vector with dimension 100 for each song's embedding. A sliding window approach is utilized the generate the embeddings for every song in the dataset. For prediction, given a prefix, this approach simply computes the nearest song to the last song in the embedding space. The baseline presented predicts the song that maximizes the cosine similarity of the embedding vector with the fifth song. Variations to this approach includes computing an aggregate embedding of the prefix by taking average of the individual song embeddings.

\subsubsection{Neural Sequence Learning}
These types of approaches aim to learn directly from the sequences using recurrent neural network architectures. Recurrent neural networks (RNN) are extremely popular in the domain of next-item prediction, mostly due to their high performance when modeling sequential data. As described by Zhu et al.zhu2017next, RNN operates on the principle that if item A has previously been seen in a sequence then items that are very similar to item A will be seen later on in the sequence \cite{zhu2017next}. 

LSTM (Long Short-Term Memory) is an example of such recurrent neural network that aims to take in a sequence of items and predict the next likely item. As part of our baseline, we utilize Keras LSTM to create an RNN model for the dataset. Due to memory constrictions, the model uses the most popular 25,000 songs as a vocabulary when building the RNN. During prediction, the model takes the first five songs of the test playlist as a seed, and computes the probability of every song in the vocabulary as the sixth song given the seed. The model then recommends the song with the highest probability. 

\subsubsection{Transformer}

Behaviour Sequence Transformer \cite{chen2019behavior} proposed by Chen et al. demonstrate the use of Transformers in a production recommender system environment. They utilize the Transformer (cite the original paper) based architecture which inherently captures the position of items across sequences and employs the self-attention mechanism. BERT (Bidirectional Encoder Represetations from Transformers) is an increasingly popular transformer-based model primarily used in the domain of natural language processing. As described by Devlin et al., BERT \cite{devlin2018bert}, when used for language, captures information from both the left and right side of a sentence. It then performs very well on tasks such as question answering and text completion. For our baseline, we treat each playlist as a sentence with each song as a "word" in the sentence. 

Our baseline utilizes DistilBERT \cite{sanh2019distilbert} that provides a smaller model compared to BERT but preserves language understanding capabilities. We provide our DistilBERT model with a vocabulary of the most popular 25,000 songs, identical to the vocabulary of the Neural Sequence Learning model proposed earlier. The DistilBERT model is evaluated using the text-completion strategy where the first five songs are given to the model as a seed and the next song is predicted by the model. 

\subsubsection{Markov Chains}

Markov Chains (MC) are one of the most popular approaches towards next-item prediction. As described by Rendle et al. \cite{rendle2010factorizing}, MC methods make item prediction by learning a transition graph over a sequence of items. This transition graph is then used to make further predictions based on the current items seen. In an n-th order Markov chain, the $n$ most recent items are used as the seed to make the prediction for the next item. For our baseline, we used n-th order Markov chain with n $\in$ \{1, 2, 3, 4, 5\}. During training, the transition graph is constructed by creating a graph of sequences $a_1, a_2,..., a_n \rightarrow a_{n+1}$. During prediction, the model recommends the graph path that occurred with highest frequency using the last $n$ songs as the sequence. In our baseline, if no sequence $a_1, a_2..., a_n$ can be found during prediction, the model will retry using the $(n-1)$-order Markov model until $n=1$. The MC approach used as a baseline is the 2-order MC. Variations to this approach include adding weights to each item in the graph sequence.

\subsubsection{Max Approach}

The max approach is a simple \& intuitive, yet a very strong baseline in our next item prediction task. Given a test instance prefix the max approach recommends the most popular item seen after the prefix in the training dataset. There are variations to the max approach which are described in the results section below.

\subsubsection{Overall Most Popular}

The overall most popular approach is a very simple baseline which merely predicts the same song, the most popular song seen in the training dataset, for every single example in the testing dataset. The most popular song was HUMBLE by Kendrick Lamar, occurring 4,608 times in the training dataset.


\section{Results}


\begin{table}[h!]
\caption{Results for the Spotify Dataset}
\centering
 \begin{tabular}{||c | c | c | c ||}
 \hline
 Approach & Accuracy & MRR@20 & Recall@20\\
 \hline\hline
 Bayesian & 8.1\%  & 0.11664 & 0.0522 \\
 \hline
 Nearest Embedding & 1.8\% & 0.0391 &  0.0277\\
 \hline
 Markov Chain & 6.8\% & 0.0937  & 0.0345\\
 \hline
 Context Max & 7.4\% & 0.10274  & 0.045975 \\
 \hline
 Overall Most Popular & 0.7\% & 0.0027 & 0.0105\\
 \hline
 SSE-PT  & 7.4\% & 0.1656 & 0.0268 \\
 \hline
 BERT4Rec & 2.7\% & 0.0773 & 0.0274 \\
 \hline
\end{tabular}
\end{table}

\vspace{50pt}

\begin{table}[h!]
\centering
\begin{tabular}{||c | c | c | c ||} 
 \hline
 Approach & Accuracy & MRR@20 & Recall@20\\
 \hline\hline
 Bayesian & 2.8\%  & 0.048 & 0.036 \\
 \hline
 Nearest Embedding & 2.1\% & 0.0429 & 0.0321 \\
 \hline
 Markov Chain & 1.9\% & 0.0317 & 0.0103\\
 \hline
 Context Max & 2.1\% & 0.0368 & 0.0419 \\
 \hline
 Overall Most Popular & 0.46\% & 0.0105 & 0.0347\\
 \hline
 SSE-PT & 1.2\% & 0.0358 & 0.0433 \\
 \hline
 BERT4Rec & 1.8\% & 0.0453 & 0.0175 \\
 \hline
\end{tabular}
\caption{Results for the Movielens dataset}
\end{table}

\section{Discussion}

Our approach is novel because we combine the best of both worlds, namely: Bayesian and Neural Network approaches. The Bayesian approach we use to model the posterior mean for a particular sequence of items represents an intuitive approach of modeling using the training data.

The neural network is used as a tool to provide an accurate estimate of the alpha and beta values used in the estimation of the posterior mean. The neural network approaches are very good black-box approximation functions that can be leveraged to analyze for patterns across large amounts of training data.

Evidently, our approach performs very well compared to multiple other baselines. Surprisingly, the seeded max approach attained an accuracy of 7.4\% which is a significant improvement from the unseeded overall most popular approach which attained an accuracy of 0.7\%. The 2-order Markov Chain produced an accuracy of 6.8\% which was the third highest accuracy. Although both neural sequence learning and transformer-based learning had relatively low accuracy, these approaches were both constrained by memory. Still, our approach is both time and space efficient while attaining excellent results.

We believe we are the first 
to leverage the benefits of both types of machine learning approaches and applied it to the domain of recommender systems.


\section{Conclusion}

In this chapter, we demonstrated a Bayesian approach to the next item prediction problem, popular in the recommender system domain. We have employed the benefits of neural networks in tasks where they are known to perform very well, which is function approximation given a large amounts of data. The main issue we tackle is that we are breaking the link between users and the items they consume on recommender system platform. This anonymity simulates the "incognito" behavior in recommender systems.

We believe our work is novel and among the first to combine the benefits of both Bayesian and statistical neural network based learning. The Bayesian approach is used to estimate the distribution instead of modeling the raw point estimate since a low frequency count does not provide a confident estimate of the underlying next-item.
Our approach alleviates the privacy concerns of existing recommender systems since we do not track a user based on their entire history of interactions. We only focus on the immediate context when deciding the next item to recommend. 

We have shown that our approach in addition to being more intuitive, also outperforms existing state-of-the-art approaches in performance. In this chapter, we have demonstrated an approach that requires users to trust the central recommender system that it will not profile the user across various sessions as currently done by existing status quo recommender systems.

\bibliographystyle{ACM-Reference-Format}
\bibliography{arxiv.bib}


\end{document}